# Sinking Satellites and Tilting Disk Galaxies

Siqin Huang and R. G. Carlberg

Department of Astronomy, University of Toronto, Toronto, Ontario, Canada M5S 1A7

## ABSTRACT

The infall of a satellite galaxy onto a galactic disk generally brings in angular momentum that is not aligned with the axis of the disk. The main dynamical issues addressed are what fraction of the orbital angular momentum of the satellite and the associated energy is added to the disk, as opposed to being left in the halo, and whether the absorbed fraction is added coherently or thermalized in the disk. By employing fully self-consistent disk+halo+satellite N-body simulations, we study the particular case of the satellite and main halo having similar density profiles, with internal velocities having the "cosmological" scaling $\sigma \propto M^{1/3}$. The satellites are introduced into the system at a distance of 10 half-mass radii of the disk and the simulations are run for about 50 disk rotations. We find that most of the orbital angular momentum of the infalling satellite is left in the tidally stripped satellite remnants, with only 2%, 6% and 9% of the orbital angular momentum being transferred to disks and halos for 10%, 20% and 30% disk-mass satellites respectively. Because the disks are tilted by the infall of 10%, 20% and 30% disk-mass satellites by angles of $(2.9 \pm 0.3)°$, $(6.3 \pm 0.1)°$ and $(9.7 \pm 0.2)°$ respectively, the kinetic energy associated with the vertical motion in the initial coordinate frame of the three disks is respectively increased by $(6 \pm 3)\%$, $(26 \pm 3)\%$ and $(51 \pm 5)\%$ whereas the corresponding disk thermal energy associated with the vertical random motion in the tilted coordinate frame is only increased by $(4 \pm 3)\%$, $(6 \pm 2)\%$ and $(10 \pm 2)\%$, respectively. The satellites cause a warp which is substantially damped over 30 disk rotations. Under our initial conditions, a satellite having up to 20% of the disk mass would produce little observable thickening whereas a 30% disk-mass satellite produces little observable thickening inside the half-mass radius of the disk but great damage beyond the half-mass radius.

*Subject headings:* Cosmology: infall rate — galaxies: dynamics, kinematics and structure

# 1 Introduction

The stellar component of a galactic disk is built up over many rotations in a time comparable to the Hubble age. The resulting disk has a vertical scale height that is about 10% of its radial scale length. The disk's vertical extent is consistent with internal heating processes alone (Lacey 1984, Carlberg 1987, Binney and Lacey 1988, Wielen 1977–although the details of vertical heating remain controversial) leaving very little room for any external source of vertical heating to be absorbed by the disk. In the past, it has been found in both numerical experiments (Walker, Mihos and Hernquist 1995, hereafter WMH, Quinn, Hernquist and Fullagar 1994, hereafter QHF, Quinn & Goodman 1986, hereafter QG) and analytical calculations (Tóth & Ostriker 1994, hereafter TO) that the infall of a 10% disk-mass satellite will cause substantial damage to a thin stellar disk. However, this result is in disaccord with the fact that in the standard CDM model, most galaxies accreted at least 10% of their mass over past 5 billion years (Bahcall & Tremaine 1988, Kauffmann, White & Guiderdoni 1993, Lacey & Cole 1993).

Binney and May (1986) have demonstrated by using straightforward test particle orbital integrations that disks are extremely resilient in the presence of a *slow* external torque: they tilt nearly as a unit and in extreme cases, they can be turned "upside down" over time. In this paper, by employing a set of fully self-consistent disk+halo+satellite N-body simulations, we show that for the case of "cosmological" satellites, thin disks are mainly tilted rather than thickened by infalling satellites. This is due to the fact that large satellite orbital angular momenta which are not aligned with disk rotational angular momenta can easily tilt the disks and tidal stripping of satellites in halos can greatly weaken the satellite's impact on disks. The spirit of this paper is to present a reasonable counterexample to the calculations which take the satellites to have much higher mass densities and are much more difficult to be tidally stripped. We do agree that dense "bowling ball" satellites are relatively immune to stripping and that once they enter the disk, tremendous vertical heating is inevitable. To achieve our end of presenting a counterexample, we have implemented in our N-body simulations several features which significantly differ from those considered in previous work (WMH, QG, QHF and TO). First, we employ similar density profiles for both the satellite and the parent galaxy. Furthermore, the initial distance between the satellite and the parent galaxy is relatively large. Finally, the disk is strongly self-gravitating which helps the disk to respond as a unit and minimizes coupling between the satellite orbit and the vertical oscillations within the disk. We believe that all of these modeling assumptions are reasonable for most of the dwarf satellites of our Galaxy.

This paper is subdivided into 8 sections. In section 2, we describe the details of our fully self-consistent N-body simulations. In section 3, we discuss the evolution of disks and satellites. We then present results on three major dynamical effects of sinking satellites on disk galaxies, namely, tilting, heating and warping. The disk tilting and heating are investigated in section 4, and the disk warping is studied in section 5. In section 6, we measure the thickness of the disk in a local tilted coordinate frame. Finally, the orbital decay and the tidal disruption of the satellites are discussed in section 7. A discussion of our main results and some concluding remarks are given in section 8.

## 2 NUMERICAL METHODS

### 2.1 Initial conditions

A scenario for the formation of disks in spiral galaxies was outlined by White and Rees (1978) and Fall and Efstathiou (1980). A dominant dissipationless background of dark matter carrying gas was assumed to cluster hierarchically. As the gas cools in the system, its pressure decreases leading to further collapse in the potential well of the dissipationless dark matter until the configuration of the disk reaches centrifugal equilibrium. We have constructed our models according to such ideas by assuming that both the halo and the satellite are two lowered isothermal spheres, King models (King 1966). Since we choose self-similar profiles for both the density $\rho_h$ of the halo and the density $\rho_s$ of the satellite, we have the relation

$$\rho_s(r'/r_s) = \rho_h(r/r_h). \tag{1}$$

In this equation, $r_s$ and $r_h$ respectively denote the radius of the satellite and halo, $r'$ ($r' < r_s$) is the distance from the center of the satellite and $r$ ($r < r_h$) is the distance from the center of the halo or the parent galaxy. When the satellite orbits inside the halo, the density of the satellite inside the tidal radius $r_{tidal}$ must be larger than its surrounding halo's density,

$$\rho_s(r', r' < r_{tidal}) > \rho_h(r). \tag{2}$$

For example, if the satellite orbits at $0.5r_h$, then the satellite particles located outside $0.5r_s$ will have been tidally stripped or, in other words, the tidal radius of the satellite is less than $0.5r_s$. The tidal stripping of the satellite doesn't depend on its density profile but depends on its distance from the parent galaxy.

As a reasonable approximation to a relaxed halo, we employ a strongly concentrated King model. The sphere is described with the dimensionless central potential $W_0 = 8.0$, which corresponds to the central concentration (King 1966),

$$c = \log(r_t/r_0) = 1.833, \tag{3}$$

in which

$$r_0 = \sqrt{\frac{9\sigma_0^2}{4\pi G\rho_0}} \tag{4}$$

is the King radius and $r_t$ is the tidal radius. In equation 4, $\rho_0$ is the central density and $\sigma_0$ is the central velocity dispersion. We employ the same central potential for the satellite sphere. Finally, the relation between mass $M$ and velocity dispersion $\sigma$ for both the satellite and the halo is taken to be that given be the spherical model of White and Frenk (1992):

$$\sigma \simeq H_0 r_0 (1+z)^{\frac{1}{2}}, \text{ or } \sigma \propto M^{1/3}. \tag{5}$$

By assuming that $\sigma = \sigma_0$ in equation 5 and employing the King model relation $M = \rho_0 r_0^3 \mu$, in which $\mu$ is a constant depending only on the dimensionless central potential $W_0$ (King 1966), we obtain from equation 4 the proportionality $r_0 \propto M^{1/3}$ between the King radius and the mass. Therefore, we have the relation $r_0 \propto \sigma \propto M^{1/3}$ which we will use to scale the halo+disk and satellite spheres.

Galactic disks are formed when the gas cools inside the halo and collapses in both the radial and vertical directions. The angular momentum of the gas halts the collapse in the radial direction, allowing only a collapse in the z direction, which then results in the formation of the disk. To mimic this effect, we have chosen one out of each eight halo particles to become a disk particle so that the resulting mass ratio of halo to disk is high enough to suppress the large scale bar instability for a very long period. All halo particles are initially sorted according to their radii from small to large radii. Each chosen halo particle initially located at spherical coordinates $(r_h(i), \theta(i), \phi(i))$ is moved to a corresponding location at cylindrical coordinates $(R(i) = r_h(i/2), \phi(i), z(i) = 0)$. Therefore, the newly formed disk is embedded inside the half-mass radius of its halo with zero thickness. The total mass ratio of halo to disk is 7 and the mass ratio inside the disk radius is 3.5. The disk particles initially only have rotation velocities, which support the disk from collapse. The rotation velocity of each particle is determined by the mass inside its radius, $v_{rot}(i) = \sqrt{GM(R(i))/R(i)}$. The post setup rotation curve of the disk and an example of a typical satellite are shown in Figure 1. The resulting disk and halo system has the initial dimensionless spin $\lambda = 0.023$, which is lower than the average value between 0.05 to 0.07 found in papers (Efstathiou & Jones 1979, Barnes and Efstathiou 1987, Warren et al. 1992). This is due to the fact that we didn't include any initial rotation in the halo.

The thickness and velocity dispersion of the disk are initially set to zero but they will increase with time as they do for real galactic disk. Such an evolution can be inferred from observations which show that there is a continuous variation of stellar kinematics from the youngest to the oldest disk stars (Wielen 1977, Carlberg et al. 1985, Edvardsson et al. 1993). For example, the relation between the velocity dispersion of disk stars and their age can be described as $\sigma \propto t^\alpha$, where $\alpha$ is approximately equal to 0.5 (Wielen 1977). Galactic disks are usually treated as quasi-equilibrium systems for a time period much shorter than the Hubble time. However, since we are studying the thickening of the disk caused by an infalling satellite in a time period comparable to the Hubble time, it is therefore necessary to consider the evolution of the disk. There are many heating sources in the disk, but it is widely accepted that the disk stars are mainly heated by scattering with giant molecular clouds (Spitzer & Schwarzschild 1953, Lacey 1984, Villumsen 1985) and spiral arms (Barbanis and Woltjer 1967, Sellwood & Carlberg 1984, Carlberg & Sellwood 1985). It is impossible to directly include scattering between stars and molecular clouds in our purely "stellar" N-body simulation. However, in N-body simulations, there is two-body relaxation which is due to the accumulation of many small deflections of the orbit of a particle arising from distant encounters with other particles. The two-body relaxation heats the disk as $\sigma \propto (T/T_r)^{0.5}$, due to the limited number of particles (Huang, Dubinski & Carlberg 1991). In the relation, $\sigma$ is velocity dispersion, $T$ is time and $T_r$ is the two-body relaxation time which is defined as (Binney & Tremaine 1987)

$$T_r = \frac{N}{8 \ln N} T_c, \tag{6}$$

in which $N$ is the number of particles and $T_c$ is the crossing time. In N-body simulation, due to the use of the softened potential $\phi_{ij} = -Gm_im_j/(r_{ij}^2 + \epsilon^2)^{1/2}$, where $\epsilon$ is the softening length, the two-body relaxation time is redefined as (Huang, Dubinski & Carlberg 1991)

$$T_r = \frac{N}{8\ln(R/\epsilon)}T_c, \qquad (7)$$

in which $R$ is the size of the system. Therefore, the heating caused by two-body relaxation can be tuned to a realistic heating rate by choosing an appropriate number of particles.

Galactic disks not only have internal heating but also have internal cooling because of the gas dissipation. For some of our simulations, this process is modeled by applying a friction only in the radial direction, $\vec{F}_{fr} = -\gamma v_R \vec{R}/R$ on the disk particles in some models. Here, $R$ is the radial coordinate, $V_R$ denotes the radial velocity and $\gamma$ is the viscosity. This friction doesn't affect the conservation of angular momentum and it also doesn't cool the disk in the vertical direction so that even in models with cooling we can still study the thickening caused by an infalling satellite. This cooling method can maintain the existence of both spiral structures and warps excited by infalling satellites for a very long period.

Normal spiral galaxies on average have one satellite in close orbit around them (Zaritsky et al. 1993). Our own Galaxy is orbited by LMC and SMC at distances of 50kpc and 63kpc from the Sun. We have chosen initial orbital parameters of satellites, such as the distances from the center of their parent galaxies, the orbital eccentricities, etc., in a way that is as realistic as possible. For example, a satellite is introduced into the system at the apocenter $r_+$ from its parent galaxy which is located at one focus of the elliptical orbit. The radius $r_+$ is chosen to be four times, or in some other cases, two and half times, that of the radius of the disk rather than equal to the radius of the disk itself as has been done in previous work (QHF). Initially, the satellites are set at a distance such that they do not penetrate the disks when they pass the disk planes. In Table 1, we list the values of the parameters associated with a series of simulations that we have investigated and for which we present results in this paper. The mass ratio of satellite to disk, $M_S/M_D$, is listed in column 2. The parameters of the satellite orbit, such as the apocenter $r_+$, the orbital inclination $\theta$ and the eccentricity $e$ are respectively listed in column 3, 4 and 5. The direction of satellite rotation $v_+/v_-$, either direct (+) or retrograde (-), is listed in column 6. The last column is the viscosity $\gamma$.

## 2.2 Numerical method

For all calculations we present in this work, we employ a "tree" N-body simulation code (Barnes & Hut 1986, Hernquist 1987, Dubinski 1988). The evolution of the system is followed by a leapfrog integrator with either hierarchical timesteps (Hernquist & Katz 1989) or constant timesteps. In the case of hierarchical timestep models, we choose the tolerance parameter $\theta = 1.0$ and the minimum timestep $\Delta t = 0.04$ time model unit which is equivalent to 1% of the disk rotation period $T_{rot} = 4$ defined at the half-mass radius of the disk. Most of our simulations were carried out for a time corresponding to 50 disk rotation periods (200 time model units or 5,000 timesteps) but some of them were carried out for as long as 60 disk rotation periods (240 Model Units). This resulted in conservation of the total energy to better than 7% over 5,000 timesteps. Simulations were also performed using fixed timestep integration with a larger timestep $\Delta t = 0.08$ and a smaller tolerance parameter $\theta = 0.7$. In this case, conservation of the total energy is satisfied to better than 3% over 2,500 timesteps. All major results presented in this paper have been confirmed with both integration schemes.

Table 1: Model Parameters

| Parameters | $M_S/M_D$ | $r_+$ | $\theta°$ | $\epsilon$ | $v_+/v_c$ | $\gamma$ |
|---|---|---|---|---|---|---|
| Model 0 | 0.0 | | | | | 0.0 |
| Model 1 | 0.1 | 4.0 | 30 | 0.2 | + | 0.0 |
| Model 2 | 0.2 | 4.0 | 30 | 0.2 | + | 0.0 |
| Model 3 | 0.3 | 4.0 | 30 | 0.2 | + | 0.0 |
| Model 4 | 0.1 | 2.5 | 30 | 0.0 | + | 0.0 |
| Model 5 | 0.1 | 2.5 | 60 | 0.0 | + | 0.0 |
| Model 6 | 0.1 | 2.5 | 90 | 0.0 | + | 0.0 |
| Model 7 | 0.1 | 4.0 | 30 | 0.0 | - | 0.0 |
| Model 8 | 0.2 | 4.0 | 30 | 0.0 | - | 0.0 |
| Model 9 | 0.3 | 4.0 | 30 | 0.0 | - | 0.0 |
| Model 10 | 0.0 | | | | | 0.025 |
| Model 11 | 0.1 | 4.0 | 30 | 0.2 | + | 0.025 |
| Model 12 | 0.2 | 4.0 | 30 | 0.2 | + | 0.025 |
| Model 13 | 0.3 | 4.0 | 30 | 0.2 | + | 0.025 |

In our simulations we employ model units with the gravitation constant $G = 1$, the total mass of the parent galaxies including the mass of the disk and the halo $M_D + M_H = 1$ and the radius of the disk or the half mass radius of the halo $R = 1$. For standard models, the number of disk particles is 10,000 and the number of halo particles is 70,000. The number of satellite particles varies between 1000 and 3000, depending on the model. In previous N-body simulation (WMH), halo particles were treated as heavier particles in order to provide better sampling of the disk and satellite components. However, this resulted in an extra heating source in the lighter particle components due to the scattering between light and heavy particles. In this paper, each particle has the same mass ensuring that there is no extra heating source in the disk. Simulations with half and twice the standard number of particles were also performed to study the effect of two-body relaxation. The softening length $\epsilon$ is usually chosen to be of order $R/N^{1/3}$, in which $R$ is the size of the system. We therefore have selected $\epsilon = 0.025$ in our simulations. A small softening length is essential to keep the disk strongly self-gravitating.

## 3  The Evolution of Galaxies and Satellites

In figure 2, we show the evolution of the disk and satellite of Model 1 projected on the $yz$ plane. Under the influence of the satellite, the outer part of the disk is tilted faster than the inner part. Therefore, a slight warp appears after T=60. As the system evolves in time, the inner part of the disk is tilted in the same plane without precessing because of the spherical halo in our model. However, the outermost part of the disk has a greater tilting angle than the inner part and it is subject to a slow precession. The projected disk looks slightly thicker than it actually is in Figure 2 because the disk is not only tilted along the x-axis but also along the y-axis due to the dynamical friction between the satellite and the disk. The thickness of the disk must be measured in the appropriate coordinate frame for each radius within the disk, which we will take to be defined by the three principal axes of inertia of each ring of material within the disk, $(x', y', z')$.

In figure 3, only the satellite particles of Model 3 are projected on the $xy$ and $yz$ planes to show the orbital decay and tidal stripping of the satellite. As the orbit of the satellite decays inside the halo, the satellite is partially tidally stripped by the halo. Once the satellite remnant enters the disk, due to the high density of the disk, it is rapidly disrupted by the tidal force of the disk, although a small undisrupted core remains in orbit around the center of the disk for as long as 35 disk rotation periods. The tidally stripped satellite particles are initially distributed around the satellite orbital plane and then evolve differently depending on their locations. The stripped satellite particles located near the disk precess around the disk axis and settle as a thick disk around the old disk due to the flatness of the gravitational potential. However, particles located far from the disk remain in the initial orbital plane for a very long period because the gravitational potential is dominated by the halo which is assumed to be spherical in our models.

In our simulations, we didn't observe strong spiral structures caused by infalling satellites because the satellites are gradually stripped as they orbit inside the halos and then rapidly stripped by the high density disks as soon as they enter the disks. Therefore, the satellites in our simulation are not dense enough to produce strong spiral structures as shown in QHF's work. However, the spiral structures in the disk can be maintained with cooling and infalling satellites. There are two conditions to preserve spiral structures in a disk. One is a cool disk environment, the other is a driving mechanism for the spiral structures. Observations suggest that driving mechanisms for spiral structures are bars or companions (Kormendy & Norman 1979). In Model 1, there is an infalling satellite, however, the spiral structures fade out very quickly because the disk is warmed up rapidly by both the spiral structures and two-body relaxation. When cooling is introduced into the disk (Model 8), the spiral structures are maintained (figure 4). In simulations with cool environment but without an infalling satellite, the spiral structures fade out faster than that observed for Model 8. We have performed a series of simulations with different parameters, such as mass ratios of halo to disk, viscosity $\gamma$, and orbital parameters of the satellite. All these parameters affect the spiral structure, warping and tilting of the disk.

## 4 DISK TILTING AND HEATING

As the orbit of an infalling satellite decays, part of the orbital energy will be transferred into the disk and halo and the rest will be kept in the satellite remnants. The orbital plane of an infalling satellite and its parent galactic disk are usually not in the same plane. The resulting dynamical issues that must be answered is what fraction of the orbital energy associated with the vertical motion of the satellite is added to the disk coherently and what fraction is thermalized in the disk. In the former case, the added energy leads to the tilting of the disk whereas in the latter case, it causes the thickening of the disk. We will calculate the total kinetic energy associated with the vertical motion of the disk particles in both the initial coordinate frame and the tilted coordinate frame so that we will be able to answer whether the disk is more likely to be tilted or thickened by an infalling satellite.

First, we present a perturbed disk in the initial coordinate frame and in the tilted coordinate frame. In figure 5, we plot the disk particles of Model 0 projected in $xy$, $xz$ and $yz$ planes and the disk particles of Model 3 projected in both $xy$, $xz$ and $yz$ planes and tilted $x'y', x'z', y'z'$ planes at T=160 time model units or 40 disk rotation periods. The $x, y, z$ are the initial coordinates and $x', y', z'$ are tilted coordinates which are determined by the three principal axes of the inertia of the tilted disk. We observe disk tilting in both the $xz$ and $yz$ planes and the disk looks much thickened in the initial coordinate frame because of the tilting, but does not appear in the tilted coordinate frame.

The gradual tilting of the disk by the infalling satellite is shown in figure 5. In figure 6, we plot the angular momentum directions of the disk, the satellite and the whole system. For a given angular momentum $\vec{L}(\theta, \phi)$, we plot its direction as a point in polar coordinates $(\theta, \phi)$. In the figure, characters D, S, T are the initial directions of the disk, satellite and total (which should, but is not exactly conserved in our simulations) angular momenta. The change in x coordinate shows that the angle between the disk plane and the satellite orbital plane decreases due to the angular momentum transfer between them. Also it can be seen from the figure that since the disk and satellite exert a torque on each other, the disk plane and the satellite orbital plane are precessing in opposite directions. We found that during 40 disk rotation periods, the directions of the disk and satellite angular momenta respectively shifted about 10° and 9°. During the same period, the drift of the direction of the total angular momentum, due to the accumulated integration error, is very small. We found it is approximately 2°, or 2″ per timestep.

In order to study the angular momentum transfer amongst the satellite, the disk and the halo, we have also measured the relative change in the magnitude and the direction of the individual angular momenta associated with these three components of the system. But we first discuss the results of a simpler case which involves the angular momentum transfer between a disk and a halo in an isolated galaxy, namely, Model 0. Since the halo is initially set without rotation, the disk rotation angular momentum is the total angular momentum of the galaxy. As the galaxy evolves with time, the disk gradually transfers its angular momentum to its surrounding halo at a steady, low rate of 0.36% per rotation period. For simulation run for a few rotation periods, the interaction between the disk and halo is small as pointed out by Sellwood (1980). However, we found that in simulations run for 50 disk rotation periods, the disk had by the end of the simulation transferred 18% of its angular momentum to its surrounding halo. Therefore the angular momentum transfer between the disk and halo is not negligible. For these simulations, the conservation of the total angular momentum is very good: there is no more than a 0.4% of variation during the 50 disk rotation periods.

When we introduce a satellite into the disk and halo system, we will additionally have to consider the angular momentum transfers between the satellite and the disk and the satellite and the halo. For a rigid or point mass satellite, the orbital angular momentum of the satellite will be completely transferred to the disk and halo system. However, for a self-consistent satellite, most of the orbital angular momentum will be kept as the rotational angular momentum of the satellite remnants due to the tidal stripping. In figure 7, we show the evolution of the satellite, disk and halo angular momenta for Model 1, Model 2 and Model 3. The left set of panels shows the evolution of the magnitudes, whereas the right set of panels shows the evolution of the directions of the various angular momenta. Only 2%, 6% and 9% of the satellites' angular momenta are transferred to disks and halos of Model 1, Model 2 and Model 3, respectively. For all three models, the magnitudes of the disk angular momenta decrease by approximately 10% because the disks transfer their angular momenta to their surrounding halos. Due to the large initial distances between the satellites and the disks, the ratios of disk angular momenta to satellite angular momenta are much larger than their corresponding mass ratios. In Model 1, Model 2 and Model 3, the angular momentum ratios respectively are 43%, 90% and 130%, whereas the corresponding mass ratios are only 10%, 20% and 30%. The large angular momenta of the satellites make disk tilting very easy. In Model 1, Model 2 and Model 3, the disks are respectively tilted by angles of $(2.9 \pm 0.3)°$, $(6.3 \pm 0.1)°$ and $(9.7 \pm 0.2)°$.

Since the dynamical friction exerted on an infalling satellite is proportional to the square of the satellite mass, a heavier satellite will be subjected to stronger dynamical friction. This stronger dynamical friction, in turn, results in greater angular momentum loss for the satellite. In the case of 10%, 20% and 30% disk-mass satellites, we find that by the end of our simulations they have respectively lost 2.2%, 5.6% and 9.2% of their angular momenta. The remaining angular momenta of those satellites are left with their remnants. As for the directions of these satellite angular momenta, they have respectively changed by angles of 7.5°, 9.4° and 9.2° relative to their initial direction in Model 1, Model 2 and Model 3. As for the halo angular momentum, it reaches a fraction approximately equal to 10% of the total angular momentum of the system for all three models, but in the heavier satellite model, the halo angular momentum is slightly larger because the halo absorbs more angular momentum from the satellite. Clearly, the halo absorbs angular momentum very efficiently both from its satellite and its embedded disk. Finally, since the initial halo was not rotating, we cannot compare the final direction of its angular momentum with its initial direction in figure 7.

The infalling satellite not only tilts the disk through the transfer of angular momentum between the satellite and the disk, but also heats the disk through the transfer of energy between these two components. In this paper, we will focus on an investigation of disk heating in the vertical direction. In figure 8, we plot the kinetic energy associated with the vertical motion of disk particles in the initial coordinate frame $k_z$, and in the tilted coordinate frame $k_{z'}$, for Model 1, Model 2 and Model 3. Here the tilted coordinate frame is constructed by employing the principal axes of inertia tensor of the disk. We found that a 10%, 20% and 30% disk-mass satellite infall increases the thermal energy associated with random motion of disk particles in the tilted coordinate frames by only $(4\pm3)$%, $(6\pm2)$% and $(10\pm2)$% respectively, as compared to the isolated model. However, in contradistinction, the kinetic energy associated with the vertical motion of disk particles is respectively increased by $(6\pm3)$%, $(26\pm3)$% and $(51\pm5)$% in the initial coordinate frames as compared to the isolated model. Our result shows that it is more likely that disks are tilted rather than heated by infalling satellites.

In order to study the distribution of the thermal energy associated with the vertical random motion of disk particles, we plot in figure 9 the evolution of the vertical velocity dispersion of the disks as a function of radius in the tilted coordinate frame. The disks are not heated uniformly as a function of radius by infalling satellites as it can be seen from the figure that both the outer and the inner regions of the disks are much more thickened than the other regions. It should not, however, be conclude that the thickening of the inner part of the disk shown in figure 9 corresponds to actual thickening. As there is warping in the disk, the tilting angles of the inner and outer parts of the disk are different from each other, and are also different from the tilting angle of the disk as measured in the tilted coordinate frame defined by the principal axes of the total inertia tensor. In fact, the direction of the axes in the tilted coordinate frame is mainly fixed by the outer part of the disk (inertia increases quadratically with distance). Therefore, the thickening of the inner part of the disk shown in figure 9 is in part artificially due to the mismatch between the inner, local, tilted coordinate frame and the one employed in our computation. To explicitly verify which fraction of the thickening of the inner part of the disk is caused by real thermal effect and which part is an artifact associated with an choice of reference frame, we will revert in the one after next section to study of the vertical velocity dispersion in local tilted coordinate frames.

The results of figure 9 can be summarized as follows: the evolution of the vertical velocity dispersion in the disk with the infall of a 10% disk-mass satellite is almost the same as the one in the isolated galaxy so there is no detectable thickening in the disk. The disk is slightly thickened by the 20% disk-mass satellite and the 30% disk-mass satellite infall definitely causes detectable thickening in the disk, especially at the outer part of the disk. Finally, in order to compare our work with previous work that suggested that a dense 10% disk-mass satellite is sufficient to thicken a thin disk, we have also shown in figure 9 that the effect that is caused by a dense 10% disk-mass satellite of half the standard size when it is introduced into the system at the edge of the disk at $r_+ = 1.0$. We find that the damage to the disk is as large as the damage that is caused by the 30% disk-mass satellite.

## 5  WARPING

Galactic warps are very common and remain a long standing puzzle in galactic dynamics. Attempts to resolve this puzzle have been summarized by Toomre (1983) and Binney (1992) and are classified into two groups: warps can either be excited during the the formation of an isolated galaxy or be excited during the interaction of galaxies. Binney (1992) suggested that the existence of warps may imply that a disk has recently been subjected to an external torque may due to an infalling satellite. Nelson and Tremaine (1995) have shown that warps are likely to be heavily damped in a dark halo so that the presence of extreme bends are unlikely to be long lived phenomena. Therefore, the fact that 50% of galaxies are warped could be an indicator that disks are in fact being subjected to the addition of considerable misaligned angular momentum. Our N-body simulations demonstrate that warps are excited by the infalling satellite (figure 2). Therefore, we will investigate in this section the kinematics of the warps. We will first construct two models: one, we will call ring model, in which case the disk is subdivided into rings and the evolution of these rings is studied. The other, we will call particle model, in which case the individual particle motions are followed.

We regroup disk particles into 10 rings (i=0-9). Each ring has its own center and rotates about its own axis. Particles located between $R(i)$ and $R(i+1)$ form Ring $i$, where the radius $R(i) = 0.2 \times 1.25^i$. The disk particles are divided in this way so that there is a sufficient number of particles in each ring, ensuring that a reliable estimate of the moment of inertia tensor is obtained. For each Ring $i$, the principal axes of its inertia tensor are used as local coordinate axes $(x'_i, y'_i, z'_i)$ which in turn are determined by three independent Eulerian angles $(\phi_i, \theta_i, \psi_i)$. In figure 10, we plot 10 rings according to their new coordinates $(x'_i, y'_i, z'_i)$ at time $T = 100$. The figure shows that the larger the radius of the ring, the more it is tilted: the warps can be seen very clearly in the figure. The number on each ring indicates the direction of the $y'_i$ axis. The outer rings, Ring 7, Ring 8 and Ring 9, show a clear warp and their directions of $y'_i$ axes show the trend of precession. We also note that the center of these rings are offset.

The evolution of the warps can be studied by following the orientation of individual rings. To describe this orientation, we employ the direction defined by the $z'_i$ axis which depends only on the two Eulerian angles $\phi_i$ and $\theta_i$. In figure 11, we show how two typical rings are gradually tilted by following the evolution of the process in terms of a string of points plotted in polar coordinates $(\theta, \phi)$. One of the rings we have selected is Ring 6 which shows a behavior of typical inner rings. To represent the behavior of typical outer rings, we have considered the evolution of Ring 9. We observe that Ring 6 is tilted with some nutation and without precession, which means that the inner part of the disk is gradually tilted in one direction without precession. In contradistinction, Ring 9 is tilted with some slow precession. This is because the ring is tilted away from the rest of disk and starts to precess under the torque imparted by its displacement. Since the precession is very slow, the warps can last for a very long time. As can be seen in figure 2, the warp caused by the satellite in Model 2 lasts for more than 30 disk-rotation periods.

An alternative way to study the warping of the disk is to follow the evolution of the direction of the orbital angular momentum of individual disk particles which are located at different radii. We have plotted in figure 12 the evolution of the directions of the angular momenta in polar coordinates $(\theta, \phi)$ for four typical particles as well as the traces of their positions. In the upper left panel. We consider the case in which a particle is located in the disk. The motion of the particle is mainly circular with little inwards shifting caused by the decreasing of rotation velocity and increasing of random velocity due to two-body relaxation. However, the direction of the angular momentum shows clear evolution. The mean value of $\theta$ increases with time, which means that the orbit of the particle is gradually tilted by the infalling satellite. An increase of the dispersion of $\theta$ with movement from the origin indicates that the particle is gradually heated due to two-body relaxation. There is no precession of the angular momentum for such a particle. In the disk, the angular momentum of most particles evolves in this way, which implies that most of the disk doesn't precess. This result is consistent with the conclusion of our ring model study, which suggests that the innermost rings do not precess. In the upper right panel, we consider the case of a particle which is very close to the edge of the disk. The particle orbits with increasingly greater radius around the center of the disk due to the increment of angular momentum caused by infalling satellite. As the particle moves outwards, its mean orbital plane is gradually tilted away from the rest of the disk because the direction of the angular momentum absorbed from the satellite is not aligned with the direction of the initial angular angular momentum. Its angular momentum starts to precess under the torque imparted by the rest of the disk. This result also is consistent with the result that we have described for the ring model, namely that the outermost rings precess. The particle shown in the lower right panel has a motion and an evolution of the angular momentum which is similar to the one depicted in the upper left panel except that the inward motion and the standard deviation of $\theta$ are larger in this case. This follows the fact that since $\theta$ is defined as $\cos^{-1}(l_z/l)$, $\theta$ becomes very large for a particle with small $l_z$ as is the case in this example of a particle located near the center of the disk. Finally, in the lower left panel, we consider a particle which is initially located at the edge of the disk. We observed that its mean orbital plane is easily tilted by the infalling satellite and that the angular momentum precesses under the torque imparted by the rest of the disk.

We conclude, from our analysis of the results obtained with both ring and particle models that the infalling satellite can excite warps in the disk. However the warps eventually fade away due to the slow precession of the outermost part of the disk.

# 6 THICKNESS OF THE DISK

Given the fact that during the evolution of the system, warping in the disk is excited by the infalling satellite, it is necessary to consider a local tilted coordinate system in order to study the disk thickening and its time dependence. We therefore measure the vertical velocity dispersion of particles in a coordinates system defined by the local direction of the angular momentum as calculated in the initial coordinate frame. To this end, we assume that the particles are distributed in a ring as shown in figure 13a. The individual direction of angular momentum $\vec{L}(\theta, \phi)$ of each particle is plotted in the terms of polar coordinates $(\theta, \phi)$ in figure 13c. In this figure, the plotted circle corresponds to a dispersion of $\theta$ of one standard deviation $\sigma_\theta$. To construct the velocity distribution of particles in the ring, we have employed a Gaussian distribution in $\theta$ and uniform distribution in $\phi$. If the ring of particles is simply tilted by an infalling satellite as shown in figure 13b, although the values of $\theta$ and $\phi$ of each particle are changed, the resulting $\sigma_\theta$ remains unchanged (Figure 13d). On the other hand, if the particles in the ring are heated by the infalling satellite, the value of $\sigma_\theta$ will increase. Therefore, we can study the evolution of the dispersion of $\sigma_\theta$ to further investigate the evolution of the vertical velocity dispersion of the particles in their local tilted coordinates without going through the rotation of coordinates. Indeed, for sufficiently small $\theta$, $\theta \approx \tan(\theta) = l_\phi/l_\theta = v_\theta/v_\phi$, in which $v_\theta$ and $v_\phi$ respectively denote the vertical and circular velocities. Since $\sigma_\theta^2 = \sigma_{v_\theta}^2/v_\phi^2 + \sigma_{v_\phi}^2 v_\theta^2/v_\phi^4 \approx \sigma_{v_\theta}^2/v_\phi^2$ follows from the fact that $v_\phi \gg v_\theta$, $\sigma_\theta \approx \sigma_{v_\theta}/v_\phi$. The dispersion in $\theta$ is therefore proportional to the vertical velocity dispersion.

In figure 14, we compare the evolution of $\sigma_\theta$ for Model 2 and 3 with that for Model 0. We didn't plot the results for Model 1 in the figure because the infall of the 10% disk-mass satellite doesn't cause significant change in the dispersion of $\theta$. We find that the infall of the 20% disk-mass satellite mainly causes heating of the disk outside of the half-mass radius which is located at 0.4 radial model unit. In the case of the infall of a 30% disk-mass satellite, the vertical velocity dispersion is increased across the disk but the outer regions of the disk are heated much more than the inner regions. Since the thickness of a disk is proportional to the squared of the vertical velocity dispersion, we can measure the increment in the thickness of the disk of Model 2 and Model 3 and compare it with the result obtained with Model 0. We find that when averaged over radii greater than the half-mass radius, the thickness of the disk of Model 2 is increased by 13%, whereas that of Model 3 is increased significantly more to 29%. For radii smaller than the half-mass radius, the average increment in the thickness of disk of Model 3 is 13% whereas that of Model 2 is only 1%. The disk is not thickened uniformly in radius, which is consistent with results of previous work (QG, QHF).

# 7 ORBITAL DECAY AND TIDAL STRIPPING OF SATELLITES

As a satellite spirals inside the dark matter halo, the distance between the satellite and its host galaxy decreases due to dynamical friction, and the mass of the satellite decreases due to tidal stripping. To investigate the orbital decay and tidal stripping of satellites, we compare our simulation results obtained for satellites with the same initial orbits but with 10%, 20%, and 30% of the disk mass (respectively Model 1, Model 2 and Model 3). We assume that the density peaks are located at the center of the satellites. If the standard deviation of the position of the density peak becomes larger than the initial half-mass radius of the satellite, the satellite is considered completely stripped. From left to right, in figure 15, we show the evolution of the distance between the satellite and the corresponding host galaxy, as well as the masses inside the initial half-mass radius of the satellite for the satellites of Model 1, Model 2 and Model 3, The mass is normalized according to the relation $f = m(r_{half})/m_{total}$ ($f(t = 0) = 0.5$). We found that during the same time period, the apocenters of the satellite orbits respectively decreased by 1.0, 2.0 and 3.0 radial model units for Model 1, Model 2 and Model 3. This is because the orbital decay rate of an infalling satellite is proportional to the mass of the satellite. Therefore, the orbit of satellites decays proportionally more rapidly as a function of increasing satellite mass. If we consider the eccentricity of a satellite's orbit, which is defined by $e = \frac{r_+ - r_-}{r_+ + r_-}$, in which $r_+$ is the orbit apocenter and $r_-$ is the pericenter, we find that for the 10% disk-mass satellite, the final $r_+$ is 3.0 and $r_-$ is 2.0, so that the final eccentricity is 0.2. For the 20% disk-mass satellite, the final orbital eccentricity is 0.18 whereas in the case of the 30% disk-mass satellite, the final satellite orbit is nearly circular. We conclude that the orbital eccentricity decreases increasingly more rapidly as a function of greater satellite mass. However, it is difficult for us to compare the results of our N-body simulations with the results obtained in analytical calculations. This is due to the complicated heating and stripping processes present in N-body simulations. Finally, the 10% disk-mass satellite is completely tidally stripped before it enter the disk. The 20% disk-mass satellite is completely tidally stripped at the edge of the disk and the 30% disk-mass satellite is rapidly stripped when it enters the dense disk.

We continue our study of the orbital decay and tidal stripping of satellites by considering satellites with 10% disk mass and an initial nearly circular orbit, but with 30°, 60°, and 90° orbital inclinations relative to the galactic plane (Model 4, Model 5, Model 6). Figure 16 shows the orbital decays as well as the change in mass inside the initial half-mass radius of the satellites. Since there is little interaction between these circular orbital satellites and the disks of their host galaxy, the orbital decay is only a weak function of the orbital inclination. The difference we note is that the satellite with a 60° inclination is completely stripped slightly earlier and further away from the host galaxy than are the satellites of the other two models. The satellites start at a radius r=2.5 and end near a radius r=1.6. The stripped satellite remnants are mainly distributed between r=1.0 and r=5.0, while the disk particles are mainly distributed between r=0 and r=1.2.

Another interesting point is to compare the orbital decay of direct (Model 1) and retrograde (Model 7) satellites. This is done in figure 17. The similar orbital decay for both satellites show that the dynamical friction exerted on the satellite by the halo is the major factor that causes the orbital decay of the satellites. The fact that the retrograde orbit decays faster than the direct one suggests that the distant interaction between the satellite and the galactic disk is not negligible although a 10% disk-mass satellite is completely disrupted in the halo before it enters the galactic disk. This is due to the fact that the Lindblad resonances of the disk exert a perturbative torque on the satellite (QG). Lynden-Bell and Kalnajs (1972) have shown that a uniformly rotating perturbing potential exerts a negative torque on inner Lindblad resonances and a positive torque on outer Lindblad resonances. In other words, the torque on the satellite due to the inner Lindblad resonances is positive and that due to the outer Lindblad resonances is negative (QG). Figure 1 shows that the outer Lindblad resonances are not present in our models, but that there are two inner Lindblad resonances. Therefore, the inner Lindblad resonances exert positive torque on the satellite. This small positive torque increases the angular momentum of a direct orbit satellite resulting in a slower decay of the satellite orbit than would otherwise occur if only dynamical friction was involved. On the other hand, we find that for the satellite moving on retrograde orbit, the torque is opposite to the angular momentum. This, as expected, decreases the angular momentum and the orbit decays faster.

For the examples considered in this section, we have shown that all 10% disk mass satellites are completely tidally stripped before they penetrate the disks. Their orbital decay is mainly caused by the dark matter halos. Therefore, the orbital parameters such as orbital inclinations of the satellite relative to the galactic disks and the direction of the satellite rotation relative to the disk rotation play little role in the satellite orbital decay.

## 8 Discussion and Conclusion

### 8.1 Discussion

We have pointed out that two-body relaxation heating is unavoidable in N-body simulations. We will first analytically estimate the velocity dispersion at a given time in the disk due to this effect, then compare the result with what is measured in our N-body simulations. Since disk particles initially only have rotation velocity, the relative velocities ($v_{rel}$) between neighboring disk particles are very small. The orbital deflections ($\Delta v_\perp^2$) of disk particles due to the distant encounters between disk and disk particles are not significant because $\Delta v_\perp^2 \propto v_{rel}^2$ (Binney and Tremaine (1987), Huang, Dubinski and Cralberg (1991)). On the other hand, the relative velocities between disk and halo particles are very large. It is therefore sufficient to consider only encounters between disk and halo particles in the following calculation of velocity dispersion of disk particles. The squared ratio of the velocity dispersion to rotation velocity of the disk particles can be expressed as a function of time $T$ and two-body relaxation time $T_r$, $\sigma^2/v_\phi^2 = T/T_r$. From equation 7, we compute the two-body relaxation time to be $T_r = 3084$ time model units by employing the following parameters: softening length $\epsilon = 0.025$, scale length of the system $R = 1$, number of particles $N = N_{halo}/2 = 35,000$ (due to the fact that the disk is initially embedded inside the half-mass radius of the halo) and crossing time $T_c = 2.6$. For a given time $T = 132$, we obtain the squared ratio of the velocity dispersion to rotation velocity $\sigma^2/v_\phi^2 = 0.042$. Since two-body relaxation heats the disk uniformly in space, we assume that the velocity dispersions in all three directions are equal, $\sigma_{v_\theta}^2 = \sigma_{v_\phi}^2 = \sigma_{v_r}^2 = 1/3\sigma^2$. We therefore find the result $\sigma_{v_\theta}^2/v_\phi^2 = 0.014$. In the above computation, we have assumed that the density distribution of halo particles is uniform. On the other hand, the density distribution of halo particles is defined by King Model in our N-body simulations. Therefore, the disk is not heated uniformly in the radial direction (can be seen in figure 14) but rather, the disk is heated more at smaller radii where the density is higher. We have selected the velocity dispersion at the half-mass radius of the disk, 0.4 radial model unit, as a typical value in the following measurements. We measured from Model 0 that $\sigma_{v_\theta}^2/v_\phi^2 = 0.015$, which is consistent with the analytical result. We have also carried simulations with half and twice the number of particles of Model 0, $N_{halo} = 35,000$ and $N_{halo} = 140,000$ respectively. The measured squared ratios of velocity dispersion to rotation velocity are 0.31 and 0.07 respectively, whereas the corresponding analytical results are 0.028 and 0.07. The good consistency between the simulation and analytical results leads us to conclude that two-body relaxation is the major source of vertical heating in our simulation and leaves little room for other sources of heating.

Therefore, the question that must be addressed is whether the number of particles we have employed to construct our models leads to a realistic source of vertical heating. As we have pointed out at the beginning of this paper, real disk galaxies have internal heating. If the two-body relaxation heating in the disks is equal or less than the real internal heating, we will be led to conclude that the simulation results are realistic. Here we compare the thickness of our Galaxy with the thickness of the disk of our simulations. Assume the vertical distribution of stars in our Galaxy is given by

$$D_S(z) = D_S(0)e^{-(|z|/\beta_S)}, \tag{8}$$

in which $\beta_S$ is the vertical scale height of stars of spectral type S. For old population dG, dK, dM stars, the scale heights are 340pc, 350pc and 350pc (Mihalas & Binney 1981).

In figure 18, we plot the vertical scale height $\beta$ as a function of time $T$ for two simulations with 80,000 and 160,000 particles. The scale height is measured at the disk half-mass radius, 0.4 radial model unit, which is 8.0kpc if we assume that the disk radius is 20kpc. Figure 18 shows that at the end of 50 disk rotation periods (200 time model units), the disk scale height for the model with 160,000 particles is less than the value measured for our Galaxy, whereas the scale height measured at the end of 50 disk rotation periods for the model with 80,000 particles is larger than the value measured for our Galaxy. In this latter case, however, the scale height measured during the interaction between the disk and the satellite, which occurs for $T < 132$ since the satellite is completely disrupted by the tidal force by the time the system has evolved through 33 disk rotation periods, is smaller than what is measured for our Galaxy. Therefore, we conclude that in our simulations, disks are sufficiently cold and thin and also are very sensitive to external vertical heating.

Both constant and hierarchical timesteps were employed in our integrations. The CPU time for a simulation with hierarchical timesteps is only 50% of the one with constant timesteps, however, the two-body relaxation heating after 50 disk rotation periods is about 10% higher in simulations with hierarchical timesteps than that in simulations with constant timesteps. Higher two-body relaxation heating in the disk yields lower disk tilting angle in response to an infalling satellite because some of the rotational energy is converted into the thermal energy. However, the disk vertical heating caused by an infalling satellite is comparable in both methods because the heating is calculated by subtracting the two-body relaxation heating from the total heating.

We have introduced cooling in some simulations to study the spiral structures. We find that cooling can effectively maintain the spiral structure in the disk. The larger the cooling rate, the stronger the spiral structures and the longer they last. We also noticed that warps caused by an infalling satellite in disks with cooling are slightly stronger and can also survive for a longer time than what are formed in disks without cooling. This derives from the fact that cooling removes some of the random motion of particles in the radial direction so that the interaction between particles at larger radii with higher inclination orbits and particles at smaller radii with lower orbital inclination decreases. Therefore warps are stronger and last longer.

Toomre (1964) investigated the stability of a thin disk to small scale and large scale disturbances. He found that small scale instabilities can be suppressed by radial velocity dispersions, $\sigma_{min} = 3.36 G\Sigma/\kappa$, or Toomre constant $Q = \sigma/\sigma_{min} = 1$. However, it is difficult to stabilize the large scale non-axisymmetric disturbances. Later studies (Hohl & Hockney 1969, Hohl 1972, Ostriker and Peebles 1973) show that this kind of bar instability can be suppressed either by decreasing the kinematic energy of rotation in the disk or embedding the disk into a dark matter halo. Therefore, it is common, in N-body simulations, to set the initial disk with Toomre constant $Q = 1$ and with a rigid halo which is two or three times disk mass. Our N-body simulations confirm that a purely rotational disk without halo or with rigid halo is unstable to bar formation as shown in early disk only simulations (Hockney 1967).

It is difficulty to completely suppress the bar instability if simulations run as long as a Hubble time since the surface density of the disk is rearranged, with more mass being placed near the center. We find that the larger the mass ratio of halo-to-disk is, the more slowly the bar develops in the disk and the wider the bar is. For example, with the mass ratio of halo-to-disk $M_h/M_d = 3.5$, a weak thick bar developed after 20 disk rotation periods. With the mass ratio of halo-to-disk $M_h/M_d = 2.5$, a narrow bar developed after 8 disk rotation periods. It is interesting to notice that the width of the bar is related to the mass ratio of halo-to-disk, though the length of the bar is determined by the rotation curve. This may explain why the observed axis ratios of bars vary among galaxies. We also find that the cooling in the disk not only can preserve spiral structures and warps in the disk as we mentioned, but also can suppress the large scale bar instability. In the simulation with $M_h/M_d = 3.5$ and $\gamma = 0.025$, the large scale bar instability can be completely suppressed during 50 disk rotation periods. However, in the simulation with $M_h/M_d = 2.5$, it requires higher cooling $\gamma = 0.1$ to suppress the large scale bar instability during 50 disk rotation periods.

We would like to emphasize the major differences in assumptions and results between TO's study and ours. First, in TO's calculation, the galactic disk is not allowed to tilt towards the direction of the orbit of the infalling satellite. Therefore, the vertical component of the orbital energy of the infalling satellite deposited in the disk is completely converted into the vertical thermal energy of the disk. Therefore, in their calculation, a thin disk is easily thickened by an infalling satellite. However, in our calculation, the angular momentum transfer between the satellite orbit and the disk makes the disk tilt towards (direct satellite orbit) or away (retrograde satellite orbit) from the initial satellite orbit plane. Thus only a small amount the orbital energy associated with the vertical motion is absorbed by the disk and is converted into vertical thermal energy in the disk. Therefore, the galactic disk is heated much less in our simulations than in TO's calculation.

TO studied disk thickening caused by satellites having density profiles described by a Jaffe Model, that is, $\rho \propto \frac{1}{r^2(1+r)^2}$. Such a functional form leads to an extremely high density at the center of the satellite as compared to the density of the parent galaxy. Therefore, the satellites could not be completely disrupted by the tidal force. For example, in low and high density cases considered by TO, they found that respectively 84% and 43% of the satellite mass cannot be tidally stripped. This is contrast to our simulations for which we find that, even in the case of the 30% disk-mass satellite, the satellites are completely disrupted before or when they enter the high density disk. However, rather than employ a Jaffe model, which may well describe high density satellites well but which is certainly not appropriate for low density satellites such as LMC and SMC, we have employed a King model for both the satellite and its parent galaxy. This implies that the core density profiles are approximately constant, in agreement with the profile $\rho \propto r^{0.0}$ inferred from the rise of the rotation curve over the core region for several dwarf spirals (Moore 1994). This is to be contrasted with the core density profile of a Jaffe model, $\rho \propto r^2$, and explains why, in our simulations, the disk is heated much less than it is in TO's calculation. However, given initial conditions similar to those selected by TO, we would expect to reproduce their result.

Whether a thin disk itself can survive a 10% disk-mass satellite infall is the question that QHF have tried to answer. We agree with them that a bare thin disk cannot survive a 10% disk-mass satellite infall. However the more realistic question we have tried to answer here is whether a thin disk embedded in a massive halo can survive a 10% disk-mass satellite infall. As we have discussed, we find that our 10% disk-mass satellite is completely tidally disrupted before it enters the disk. Most of its angular momentum and energy is carried by its remnants whereas the remaining small fraction of the angular momentum and energy is primarily transferred to the halo. Therefore, with the protection of a self-consistent halo, we conclude that a thin self-gravitating disk can survive a 10% disk-mass satellite infall without an observable trace of disturbance to its vertical structure.

## 8.2 *Conclusions*

In contrast to previous work (TO, QHF, WMH) in which it was suggested that a thin disk could be significantly damaged by a 10% disk-mass satellite infall, we have presented a counterexample, defined by a lower density satellite, larger initial distance between the satellite and the parent galaxy and a strongly self-gravitating disk, which demonstrates that a thin disk can survive a 10% disk-mass satellite infall. In our models, all three components, disk, halo and satellite, are fully self-consistent and the densities of the satellite and its parent galaxy as well as their initial relative distance are scaled cosmologically. Our "cosmological" satellite can be substantially disrupted by tidal forces, roughly in proportion to the fraction of the host's mass traversed as it spirals inward. Since most of the satellite's orbital energy is left in the satellite remnants, the energy which is available to be transferred into the halo and the disk is only a small amount. Also, since our self-consistent halo is very efficient in absorbing the satellite's orbital energy, the damage to the thin disk due to the infalling satellite is greatly diminished compared to the damage that occurs in the case of a rigid halo model (QHF). Although, WMH employed a self-consistent halo model, the orbital decay of the satellite in their model is mainly caused by the interaction with the disk due to the small initial distance between the disk and the satellite. Therefore, WMH's self-consistent halo does not protect the thin disk against being thickened by the infalling satellite. Another characteristic of our simulation is the large initial orbital angular momentum of the satellites which results from initially locating satellites at large distances from the disks. Therefore, the disks are easily tilted towards, or away, from the satellite orbital plane; indeed, we find that our thin disks primarily respond to an infalling satellite by tilting. Thus, disk heating in the vertical direction, or disk thickening in our models is decreased relative to TO's models in which disk tilting is not allowed and QHF's and WMH's models in which disk tilting angles are very small as the initial orbital angular momenta of their satellites are small. We have found that a 10% disk-mass satellite is completely disrupted by the tidal force before it enters the galactic disk. Therefore, the damage of the infalling satellites on the disk is not observable. Also, we have shown that a 20% disk-mass satellite is completely tidally disrupted at the edge of the disk and the disk is mainly heated outside the half-mass radius. The thickness increments, averaged inside and outside the half-mass radius of the disk, are 1% and 13% respectively. Finally, we have found that a 30% disk-mass satellite can only survive shortly in the dense galactic disk. The thickness increments, averaged inside and outside the half-mass radius of the disk, are 13% and 29% respectively.

We would like to thank Scott Tremaine and Simon Lilly for useful discussions and suggestions and Giovanni Pari for useful discussions and carefully reading the draft. HPC High Performance Computing Center in Calgary, Canada provided us some CPU time in the form of Fujitsu Scholarship, which is very appreciated. Siqin Huang would like to acknowledge the financial support from the Department of Astronomy and University of Toronto.

# FIGURE CAPTIONS

FIG. 1.— Initial conditions imposed on the disk and satellite. The initial position and orbit of the satellite are shown in the left panel. On the right panel, the dots represent the measured angular velocities of the disk and the stars are measured angular velocities of the satellite at different positions. The solid line represents the best fitting result.

FIG. 2.— Evolution of the disk and satellite particles of Model 2 as viewed in the initial disk plane ($yz$ plane). The disk particles and the satellite particles are distinguished by big dots and small dots respectively. In the figure, the dominant halo particles are not plotted. The rotation period at half-mass radius (hereafter the disk rotation period) is 4. Note the slight warp for times $T > 60$. The size of the boxes is 6 radial model units.

FIG. 3.— Evolution of the satellite particles of Model 2 projected in both the $xy$ (first two rows) and $yz$ (last two rows) planes. Halo particles and disk particles are not shown. The size of the boxes is the same as that in figure 2.

FIG. 4.— Evolution of the disk particles of Model 8 projected in the $x'y'$ plane. The dashed lines represent the $x'$ and $y'$ axes. The spiral structure is still very clear at the end of the simulation. The size of the boxes is 1/3 of the one in figure 2.

FIG. 5.— Disk particles of Model 0 projected in $xy$, $xz$ and $yz$ planes and disk particles of Model 3 projected in both $xy$, $xz$ and $yz$ planes and tilted $x'y', x'z', y'z'$ planes at T=160 time model units or 40 disk rotation periods.

FIG. 6.— Evolution of the disk (D), satellite (S) and total (T) angular momentum directions. In 40 disk rotation periods, the changes in the disk and satellite angular momentum directions are about 10° and 9° respectively, and the drift of the direction of the total angular momentum is about 2°.

FIG. 7.— Angular momentum transfer between satellite, disk, and halo as a function of time. From top to bottom, we show the evolution of the angular momenta of satellite, disk and halo in Model 1, Model 2 and Model 3 respectively. Most of the satellite initial angular momentum remains in the satellite remnants. The initial rotationless halo is very efficient in absorbing angular momentum from both the satellite and disk.

FIG. 8.— Kinetic energy associated with the vertical motion in the initial coordinate frame $k_z$ and in the tilted coordinate frame $k_{z'}$ as a function of time for, from top to bottom, Model 1, Model 2 and Model 3. After a satellite which has 10%, 20% or 30% disk mass falls into the disk, the kinetic energy associated with the vertical random motion of the disk $k_{z'}$ is respectively increased by only $4\pm3\%$, $6\pm2\%$ or $10\pm2\%$ as compared to an isolated galaxy.

FIG. 9.— Evolution of the vertical velocity dispersion as a function of time for disks with 10%, 20% and 30% disk-mass satellite infall. In each panel, the set of four pairs of curves represents, from the bottom to top, the distribution of the vertical velocity dispersion at T=0, 60, 120 and 180 time model units. Only the 30% disk-mass satellite infall, or the dense 10% disk-mass satellite infall, can cause detectable thickening in the outer part of the disks.

FIG. 10.– The 10 rings of Model 3 projected on the $xz$ plane.

FIG. 11.– Evolution of the two Eulerian angles $(\phi, \theta)$ plotted in the polar coordinates frame in degree of Ring 6 and Ring 9. Ring 6 is only tilted with some nutation, but without precession. However Ring 9 precesses very slowly.

FIG. 12.– Evolution of the angular momentum directions $(\theta°, \phi°)$ plotted in polar coordinate frame for four typical disk particles. The traces of their positions $(r, \phi')$ also plotted in polar coordinate frame which are shown in the upper right corner of each panel. The symbol $\star$ indicates the initial positions and angular momentum directions of the particles. The simulation is run for 50 rotation periods.

FIG. 13.– Particles located on the original ring (a) and the tilted ring (b). The directions, in polar coordinates $(\theta, \phi)$, of the angular momenta $\vec{L}$, of these particles are plotted in figure c and d.

FIG. 14.– Evolution of the dispersion of $\theta$ in Model 2 and Model 3 in comparison with an isolated galaxy. In each panel, we plot the distribution of the dispersion of $\theta$ at T=0, 60 and 120 time model units. The disk is mainly heated in its outer regions which is accord with the result presented in figure 9.

FIG. 15.– Orbital decay and tidal stripping of the satellite in Model 1, Model 2 and Model 3. We plot the distances between the satellites and their parent galaxies in x-axis, and the masses inside the half mass radius of the satellites in the y-axis. The heavier the satellite, the faster the orbit decays and the faster the orbital eccentricity decreases. The 10% disk-mass satellite is completely tidally stripped before it enters the disk. The 20% disk-mass satellite is completely tidally stripped at the edge of the disk and the 30% disk-mass satellite is rapidly stripped when it enters the dense disk.

FIG. 16.– Orbital decay and mass stripping of satellites with 30°, 60° and 90° orbital inclinations. Column 1 shows that the ratio of the mass inside the initial half-mass radius to the total mass of satellite (f, f(t=0)=0.5) decreases as the distance between the satellite and the galaxy (r, r(t=0)=2.5) decreases. Column 2 shows how the distance r changes with time t. Column 3 shows the mass ratio f versus time t.

FIG. 17.– Orbital decay of the direct and retrograde orbital satellites as a function of time. The retrograde orbit decays faster than the direct one.

FIG. 18.– Scale height as a function of time. The scale height for 160,000 particle model at the end of simulation is less than what is measured in our Galaxy whereas the scale height for 80,000 particle model before T=132 when the satellite is completely tidally disrupted is less than the value measured in our Galaxy. Therefore, the disks in our simulations are cold, thin and sensitive to external vertical heating.

23